\documentclass[english]{article}
\usepackage[utf8]{inputenc}
\usepackage[T1]{fontenc}
\usepackage{babel}
\usepackage{amsmath}
\usepackage{graphicx}
\usepackage{fancyhdr}
\usepackage{color}
\usepackage{mathtools}
\usepackage{listings}
\usepackage{hyperref}
\usepackage[super]{natbib}

\newcommand{\td}[1]{}
\renewcommand{\td}[1]{{\color{red} TODO: {#1}}}
\newcommand{\angstrom}{\mbox{\normalfont\AA{}}}
\def\bm#1{\mathbf{#1}}

\begin{document}

\title{A representation-independent electronic charge density database for crystalline materials}

\author{Jimmy-Xuan Shen\textsuperscript{1},
	Jason M. Munro\textsuperscript{2},
	Matthew K. Horton{\textsuperscript{1,3}},\\
	Patrick Huck\textsuperscript{2},
	Shyam Dwaraknath\textsuperscript{3}
	Kristin A. Persson\textsuperscript{1,3,*}
}

\maketitle
\thispagestyle{fancy}

1. Department of Materials Science and Engineering, University of California, Berkeley, Berkeley, California 94720, United States

2. Energy Technologies Area, Lawrence Berkeley  National Laboratory, Berkeley, 94720, United States

3. Energy Sciences Area, Lawrence Berkeley  National Laboratory, Berkeley, 94720, United States

	{*}corresponding author(s): Kristin A. Persson (kapersson@lbl.gov)

\begin{abstract}
	In addition to being the core quantity in density functional theory, the charge density can be used in many tertiary analyses in materials sciences from bonding to assigning charge to specific atoms.
	The charge density is data-rich since it contains information about all the electrons in the system.
	With increasing utilization of machine-learning tools in materials sciences, a data-rich object like the charge density can be utilized in a wide range of applications.
	The database presented here provides a modern and user-friendly interface for a large and continuously updated collection of charge densities as part of the Materials Project.
	In addition to the charge density data, we provide the theory and code for changing the representation of the charge density which should enable more advanced machine-learning studies for the broader community.
\end{abstract}

\section*{Background \& Summary}

The application of Density Functional Theory (DFT) to many-electron systems has witnessed tremendous growth in the past few decades and has now become the {\it de facto} simulation tool for physicists, chemists, and materials scientists.
The central concept of DFT is that the energy, and in turn all of the physical properties of a quantum system, are completely determined by the electronic charge density of the ground state $\rho(\mathbf{r})$ with $\mathbf{r}$ the position vector~\cite{hohenberg_kohn_1964}.
The majority of the computational cost in typical DFT calculations is associated  with determining $\rho$ via an iterative algorithm to arrive at a self-consistent charge density~\cite{kohn_sham_1965}.
For the most commonly used exchange-correlation functionals, like the local density approximation (LDA)~\cite{kohn_sham_1965,ceperley_ground_1980} and the semi-local functional by Perdew–Burke-Ernzerhof (PBE)~\cite{pbe}, a converged charge density can be used as the starting point for more expensive calculations such as obtaining a detailed bandstructure~\cite{Martin2004} or calculating the optical response of the material~\cite{gajdos_linear_2006}.

% \td{If we want to advertise that you can restart the calculations we need to make sure that the PAW components are written in pymatgen.io.vasp.Chgcar}
% Actually this might be kind of pointless since it only allows them to repeat our calculation verbitim which is wasteful
% If we can figure out how the PAW's transformations charge under a symmetry operation on a grid that would be cool but that is way more work than then entire rest of this paper. 

In addition to its central role in standard DFT calculations, the charge density itself is also useful for the analysis of many materials properties.
The critical points of the charge density (i.e. where the gradient is zero) are often used as a boundary between atomic neighborhoods. In turn, this allows for a systematic assignment of charge to specific atoms~\cite{Bader1994Jun,Popelier2001Apr}, as well as the determination of bonding character between neighboring pairs~\cite{Otero-de-la-Roza2014Mar}.
Within the realm of energy materials, the charge density can be used as an effective potential to study the migration properties of Li in solid-state materials, as low charge density provides a metric of ``free'' space in a lattice~\cite{Rong2016Aug,Kahle2018Jun}. Consequently, the local minima of the charge density can act as initial guesses for the positions of inserted cations~\cite{Shen2020Oct}.

A single DFT calculation of the primitive unit cell provides one representation of the charge density within that particular basis set. However, depending on the data application, alternative representations might be desired.
An important example of this is in machine-learning (ML) algorithms where obtaining a consistent data representation is absolutely essential for deep-learning methods.
However, the representation of charge density is heavily influenced by the simulation cell and the Bravais lattice of the periodic structure.
Hence, a necessary step in using electronic charge densities in machine-learning applications is the ability to obtain alternative representations of the same physical density.
While recent work has examined the effectiveness of representations in Fourier space~\cite{Kajita2017Dec}, any ML investigation of local interaction (e.g. adsorption and intercallation of ions) requires flexible representations in real space.
Towards that end, our database will provide code to obtain arbitrary real space representations of charge density for a given material directly from a DFT-computed charge density.

% Modern implementations of DFT are typically specialized either for the simulation of molecules or the simulation of crystalline solids.
% The specialization manifests in the choice of the basis set used to represent the system.
% For the molecular simulations, it is most natural to work with an atomically-centered basis-set, where the charge density is represented as a vector of projection values onto the basis-set. \td{Maybe give examples of molecular databases}
The charge density of any crystalline solid, and indeed any periodic field, is naturally represented in a plane-wave basis set, where the inherent periodicity of the system is embedded in the underlying representation.
For a sufficiently converged finite plane-wave basis, the continuous charge density $\rho({\bf r})$ and its Fourier transform $\phi({\bf k})$ can be accurately sampled by a three-dimensional array indexed by $i$, $j$, and $k$ with $N_1$, $N_2$, and $N_3$ evenly spaced grid-points along each lattice vector, and can be converted from one to the other via a discrete Fourier transform
\begin{align}\label{eq:dft}
	\rho(\bm{r})
	\equiv
	(\bm{a}_1,\bm{a}_2,\bm{a}_3, \rho_{i,j,k})
	\xLeftrightarrow[\hspace{1em}\mathcal{F}^{-1}\hspace{1em}]{\mathcal{F}}
	(\bm{b}_1,\bm{b}_2,\bm{b}_3, \phi_{i,j,k})
	\equiv
	\phi(\bm{k})\,.
\end{align}
where the $\bm{a}_\alpha$ and $\bm{b}_\alpha$ represent the real and reciprocal lattice vectors, and $i$, $j$, and $k$ are the indices of regularly-spaced grid points along the lattice vectors.
Due to the discrete nature of numerical Fourier transforms, the number of grid points of a real-space representation is equal to the number of plane waves needed to represent the data in reciprocal space.

A representation of the charge density is uniquely determined by three vectors and a scalar matrix either in real or reciprocal space.
Each representation only provides a ``view'' of the infinite periodic data represented in a specific unit cell and an infinite number of such representations exist for a given charge density.
Regardless of the grid size and the periodic cell representation, the DFT-computed charge density represents the same underlying field, yet they are routinely recomputed when any change is needed in the representation, even when the computational parameters are unchanged, often at considerable expense.
One common example is the use of the electrostatic potential of a super cell to correct for the periodic image effects of a charged defect~\cite{freysoldt2014}.

Due to the significant amount of computational resources devoted to computing the electronic charge densities, as well as the growing domains of their application especially for the training of data-intensive machine learning models, there is a pressing need for a large-scale representation-independent database of charge densities.
The Materials Project (\url{https://materialsproject.org}) - as a rapidly growing (currently more than 170,000 users) materials informatics resource - is a  natural platform for the dissemination of such data.
The work presented in this article provides details on how the charge densities in our database are computed and how they can be accessed.
In addition, we provide a high-level API for querying and post-processing of the charge density data.
Among other features, the API will allow users to take an existing atomic structure and query for charge density of the same material, in the representation/view of the user's choosing.

\section*{Methods}

In this section, we provide details on the scope of the charge densities database (CDD) and the precise set of computational parameters used to generate the data.
Additionally, we will demonstrate features of the API that allow users to obtain arbitrary views of the charge density data including up-sampling/compressing the data via Fourier analysis and symmetry operations like translations, rotations, and super-cell transformations.

\subsection*{Calculation parameters}

The charge densities are obtained from DFT calculations run using the static calculation workflow within the atomate software package~\cite{Mathew2017Nov}, and relaxed input structures from the Materials Project (MP) database~\cite{jain2013}.
The projector-augmented wave (PAW) method as implemented in the plane-wave Vienna Ab-initio Simulation Package (VASP) is used in conjunction with the PBE generalized-gradient approximation functional~\cite{pbe}.
The default set of MP calculation input parameters was used, which have been demonstrated to produce well-converged results~\cite{Jain2011Jul}.
Included in these parameters is an energy cutoff of 520~eV, a total energy error threshold of $5~\times~10^{-5}$~eV/atom, and a reciprocal $k$-point density of $100/\mathrm{A}^{-3}$.
The only addition made to the input set is to enable aspherical contributions in the gradient corrections inside the PAW spheres.
Hubbard $U$-corrections are included with materials containing oxygen and fluorine. Elements Co, Cr, Fe, Mn, Mo, Ni, V, and We use values of 3.32, 3.70, 5.30, 3.90, 4.38, 6.20, 3.25, and 6.20~eV, respectively.

\subsection*{Changing the charge density representations}

\begin{figure}[!htbp]
	\centering
	\includegraphics[width=0.95\textwidth]{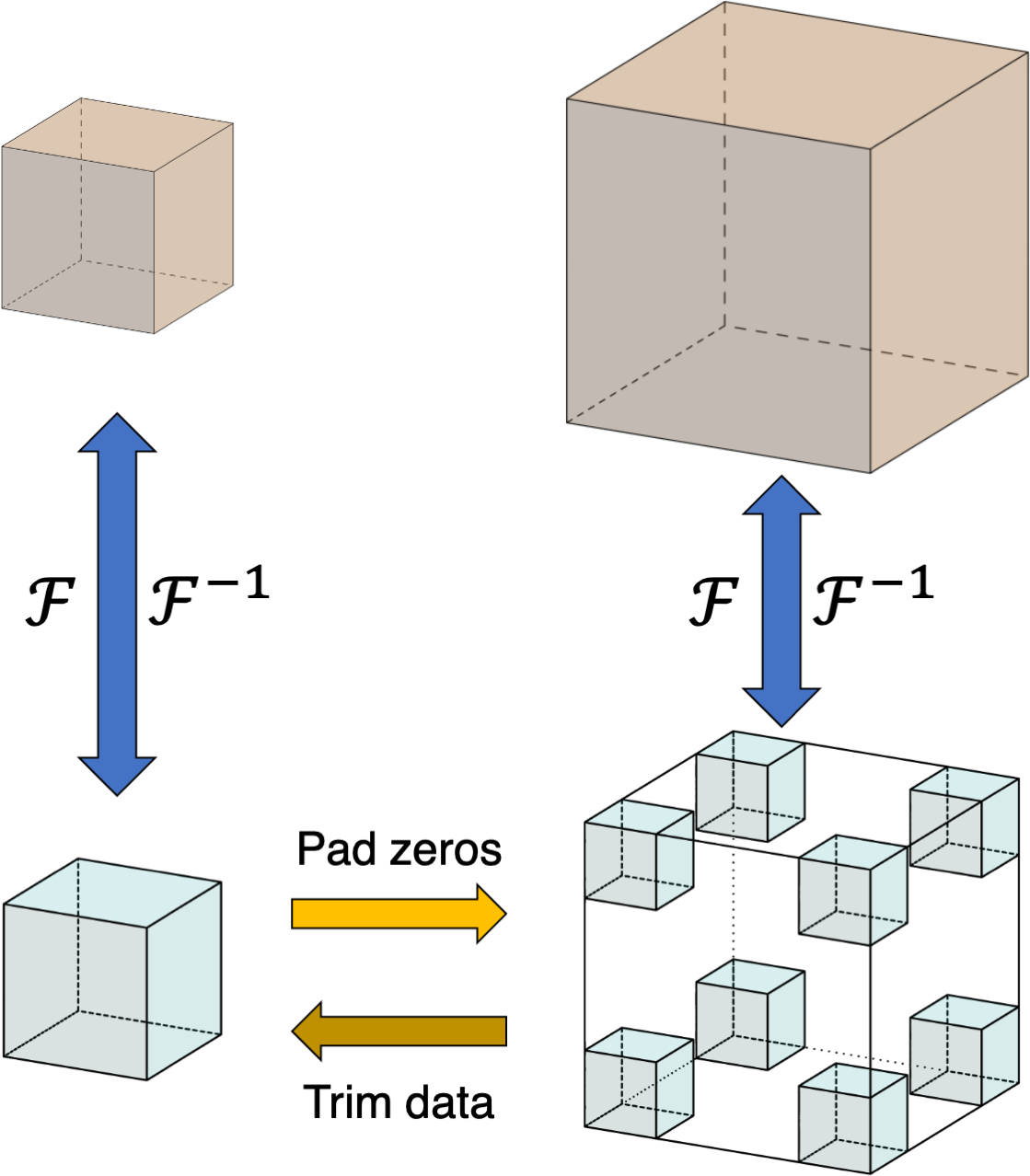}
	\caption{
		Schematic of data transfer for Fourier interpolation and compression.
		The more densely sampled (larger) and the coarsely sampled (smaller) 3D blocks of real-space data can each be transformed to Fourier space, resulting in a Fourier representation of the same size.
		To up-sample the data, we use the smaller block in Fourier space, augment with zeros while keeping all the data fixed near the origin (at the corners of the cube). 		To compress, we crop the data in Fourier space and perform an inverse Fourier transform.
	}
	\label{fig:fft_interp}
\end{figure}

Given one representation $(\bm{a}_1, \bm{a}_2, \bm{a}_3, \rho_{i,j,k})$ of the charge density $\rho$, we may transform it to any other representation $({\bm a}_1^\prime, {\bm a}_2^\prime, {\bm a}_3^\prime, \rho_{i^\prime,j^\prime,k^\prime})$ by interpolating the data.
Due to computation time and data storage constraints, DFT codes will typically use the fewest grid points possible to represent the charge density which limits the effectiveness of local interpolation schemes.
% Thus, the data quality can degrade after any resampling operation like any form of local interpolation.
However, since our charge densities have periodic boundary conditions and are reasonably smooth (owing to the use of pseudo-potentials), the charge density can be represented in Fourier space.  we can up-sample our data via Fourier interpolations~\cite{Russell2015Feb} as shown in Figure~\ref{fig:fft_interp}.
The procedure to perform Fourier interpolation of real space data is as follows:
\begin{enumerate}
	\item Take the discrete Fourier transform of $\rho_{i,j,k}$.
	\item Augment the resulting Fourier data $\phi_{i,j,k}$ with zero-valued higher frequency components.
	\item Apply the reverse transformation to obtain the up-sampled data.
\end{enumerate}
The augmented Fourier data is mathematically equivalent to the original Fourier data.
Thus, the inverse transform of the augmented Fourier data must be equivalent to the original real space data sampled at a higher density. Increasing the grid density using Fourier interpolation enables us to up-sample $\rho_{i,j,k}$ in each direction by a factor of $\gamma_{\rm up}$. We may then resample the local grid with a linear interpolation scheme to ensure the fidelity of our data.

Given a primitive-cell representation of the charge density ---
$(\bm{a}_1,\bm{a}_2,\bm{a}_3,\rho_{i,j,k})$,
any periodic representation of a scalar field $f(\bm{r})$ can be understood as applying an arbitrary translation on the unit cell by a vector $\bm{t}$:
\begin{align}
	\hat{T}_{\mathbf{t}} f(\bm{r}) \equiv f(\bm{r} - \bm{t}) \, ,
\end{align}
followed by a super-cell transformation $\hat{P}$ defined as an integer matrix with $\det(\hat{P}) \ge 1$ which acts on the lattice vectors from the right
\begin{align}
	(\bm{a}^\prime_1 \, \bm{a}^\prime_2 \, \bm{a}^\prime_3)
	=
	(\bm{a}_1 \, \bm{a}_2 \, \bm{a}_3)\hat{P} \, .
\end{align}

\begin{figure}[!htbp]
	\centering
	\includegraphics[width=0.90\textwidth]{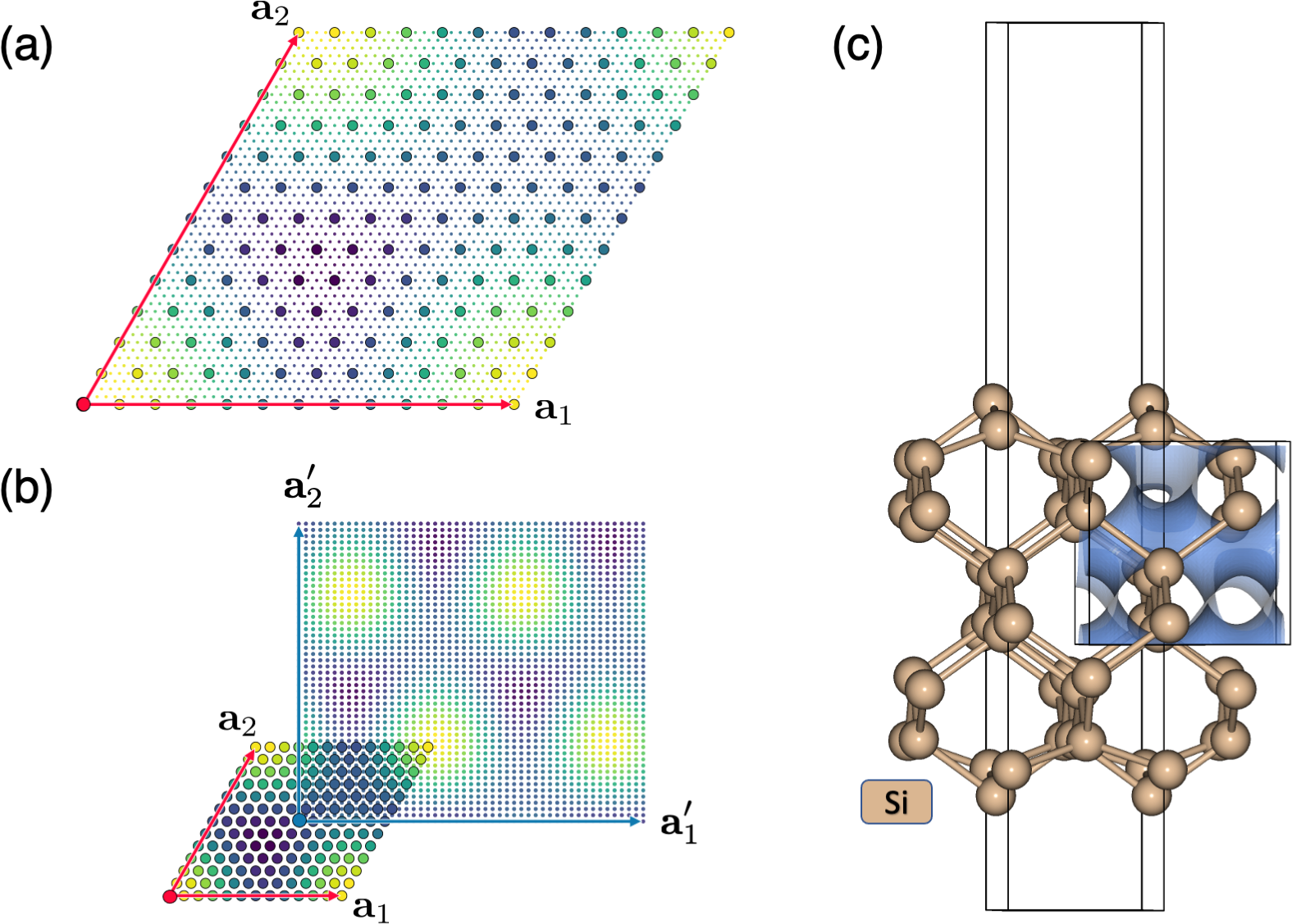}
	\caption{
	Periodic re-gridding applied to a plane of the charge density for a two atom Si unit cell where each large circle corresponds to a data point in the original $12\times12$ grid.
	The results of Fourier-interpolating the data in the original unit cell onto a $48\times48$ grid is shown in (a). The transformed representation $\hat{\bm a}_1 = {\bm a}_1 + {\bm a}_2$ and $\hat{\bm a}_2 = 2{\bm a}_1 - {\bm a}_2$ with a shift of $0.4 \hat{\bm a}_1 $ and a grid size of $48\times48$ is shown in (b).
	}
	\label{fig:resampling}
\end{figure}

Our software is capable of performing the same operations in arbitrary dimensions. As an example, in Figure~\ref{fig:resampling}, we show the results of re-griding in using a plane of the charge density in a two-atom Si unit cell which only cuts across a single atom at the origin,
Figure~\ref{fig:resampling} (a) shows the result of Fourier interpolating the field from a $12\times12$ grid (large circles) onto a $48\times48$ grid (smaller circles).
In Figure~\ref{fig:resampling} (b), the modified representation is obtained by first shifting the origin to the center of the cell at $\bm{t} = ({\bm a}_1 + {\bm a}_2)/2$ followed by a change of basis to ${\bm a}^\prime_1 = 2{\bm a}_1$ and ${\bm a}^\prime_2 = 2{\bm a}_2 - {\bm a}_1$.

While integer-valued supercell transformations will yield an equivalent periodic charge density, non-integer basis transformations are used to obtain an arbitrary crop of periodic charge density sampled at any density.
As an example, we show how a non-periodic cubic sample of the surface charge density can be obtained from the slab calculation in Figure~\ref{fig:resampling} (c).
The simulation was performed using a $7.73~\angstrom\times 3.87~\angstrom \times 21.88~\angstrom$ orthorhombic Si slab cell and the charge density is stored on a $120 \times 60 \times 336$ grid.
A $5~\angstrom \times 5~\angstrom \times 5~\angstrom$ cropped region of the charge density sampled on a $48\times48\times48$ grid is indicated by the blue iso-surface in Figure~\ref{fig:resampling} (c).
It is important to note that the cropped cell can be greater in any dimension, as compared to the simulation cell.  In the example, the smallest dimension of the simulation cell is 3.87~\AA{} while the cropped cube has side lengths of 5~\AA{}. This feature essentially allows us to robustly obtain the charge density in any preferred real-space dimensions,  independent  of the simulation cell parameters.

essentially allows us to freely choose the simulation cell in situations where periodic-image effects are not present.

\subsection*{Database details}

\begin{figure}[!htbp]
	\centering
	\includegraphics[width=1.0\textwidth]{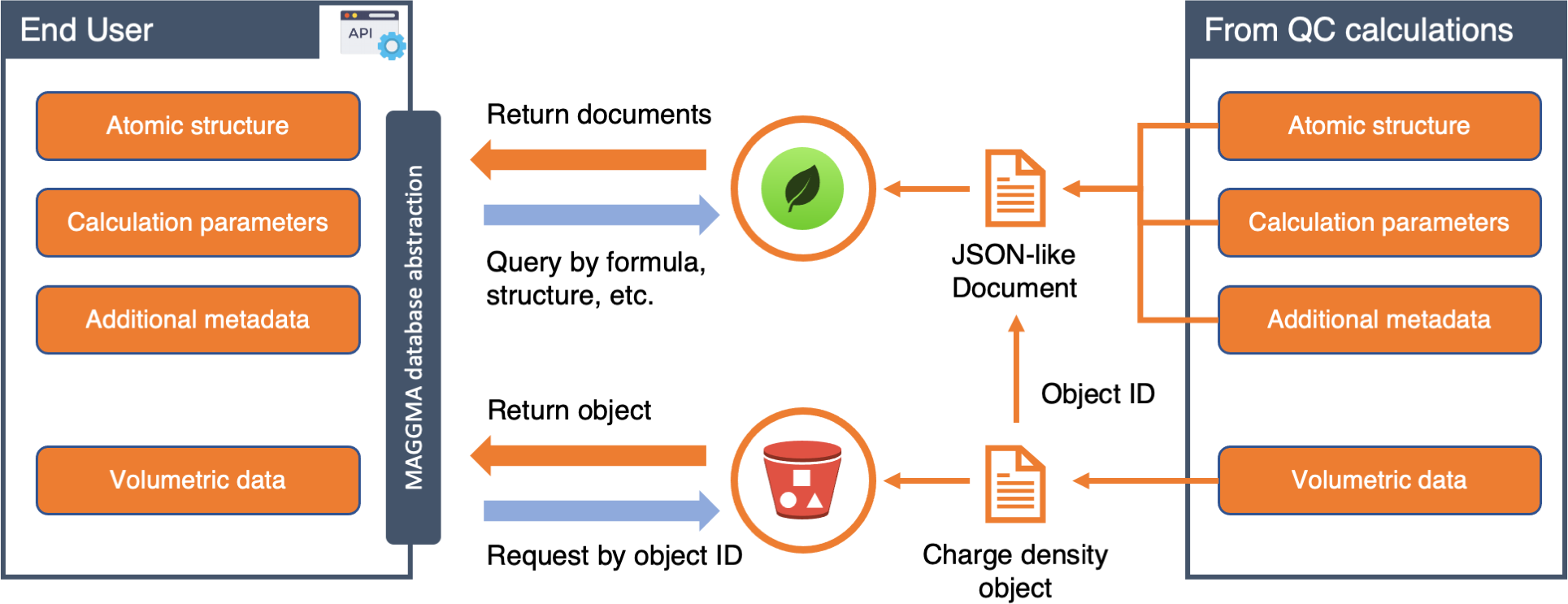}
	\caption{
		Data pipeline for the charge density database, illustrating how the output files from quantum chemistry calculations are stored and accessed. 		The data are first converted into a JSON-like format to be stored on a MongoDB server, which allows queries on any of the stored fields.
		The charge density data is converted into an array-like object with additional meta-data (e.g. \texttt{ObjectID}) and stored in an AWS S3 bucket.
		Since the \texttt{ObjectID} is stored as a field in the MongoDB database, the API is able to combine the MongoDB data with the according S3 object and reconstruct the original data.
	}
	\label{fig:flowchart}
\end{figure}

We use a hybrid data model to serve the data: Queryable data such as chemical formula, total energy, and calculation parameters are served as JSON-like documents using MongoDB, while much larger and not-queryable charge density data is served using AWS S3 object storage~\cite{Leeper2017Jun}.
When a charge density is parsed from the output file to a serialized object, a unique Object ID is assigned and stored alongside the other data in the MongoDB database.
From the user's perspective, two subsequent API requests are needed. One to obtain calculation inputs and outputs from MongoDB, and another for  Object ID and charge density data.
A visual representation of the data flow is provided in Figure~\ref{fig:flowchart}.

\subsection*{Code availability}

The software used to access and transform the charge density data is accessible from the Materials Project API (\url{https://github.com/materialsproject/api}) and \texttt{pyRho} (\url{https://github.com/materialsproject/pyRho}) python package repositories on Github. See the \hyperref[sec:usage]{Usage Notes} section for more information.

\section*{Data Records}

Raw charge density data output from DFT calculations can be obtained from the corresponding MP API endpoint:
\url{https://api.materialsproject.org/charge_density}. Each entry can be referenced with a particular DOI through the associated MP material. Additionally, the input parameters for the specific calculation used to generate the entry can be obtained from the tasks endpoint at  \url{https://api.materialsproject.org/tasks}. Details for how to interact with the referenced endpoints can be found in the \hyperref[sec:usage]{Usage Notes} section.

\section*{Technical Validation}

We can elucidate the performance of the re-griding algorithm using a larger set of elemental polymorphs from the Materials Project.
For this test set $\mathcal{S}_{\rm el}$, we selected 389 single-element structures from MP for which the energy above the convex hull was less than 100~meV and the number of atoms in the unit-cell was less than 20.
For each structure in $\mathcal{S}_{\rm el}$, we perform VASP static calculations on the primitive unit cell and on a super-cell using
\begin{align}
	\hat{P} =
	\begin{pmatrix}
		1 & 1  & 0 \\
		1 & -1 & 0 \\
		0 & 0  & 1
	\end{pmatrix} \, .
\end{align}

\begin{figure}[!htbp]
	\centering
	\includegraphics[width=0.7\textwidth]{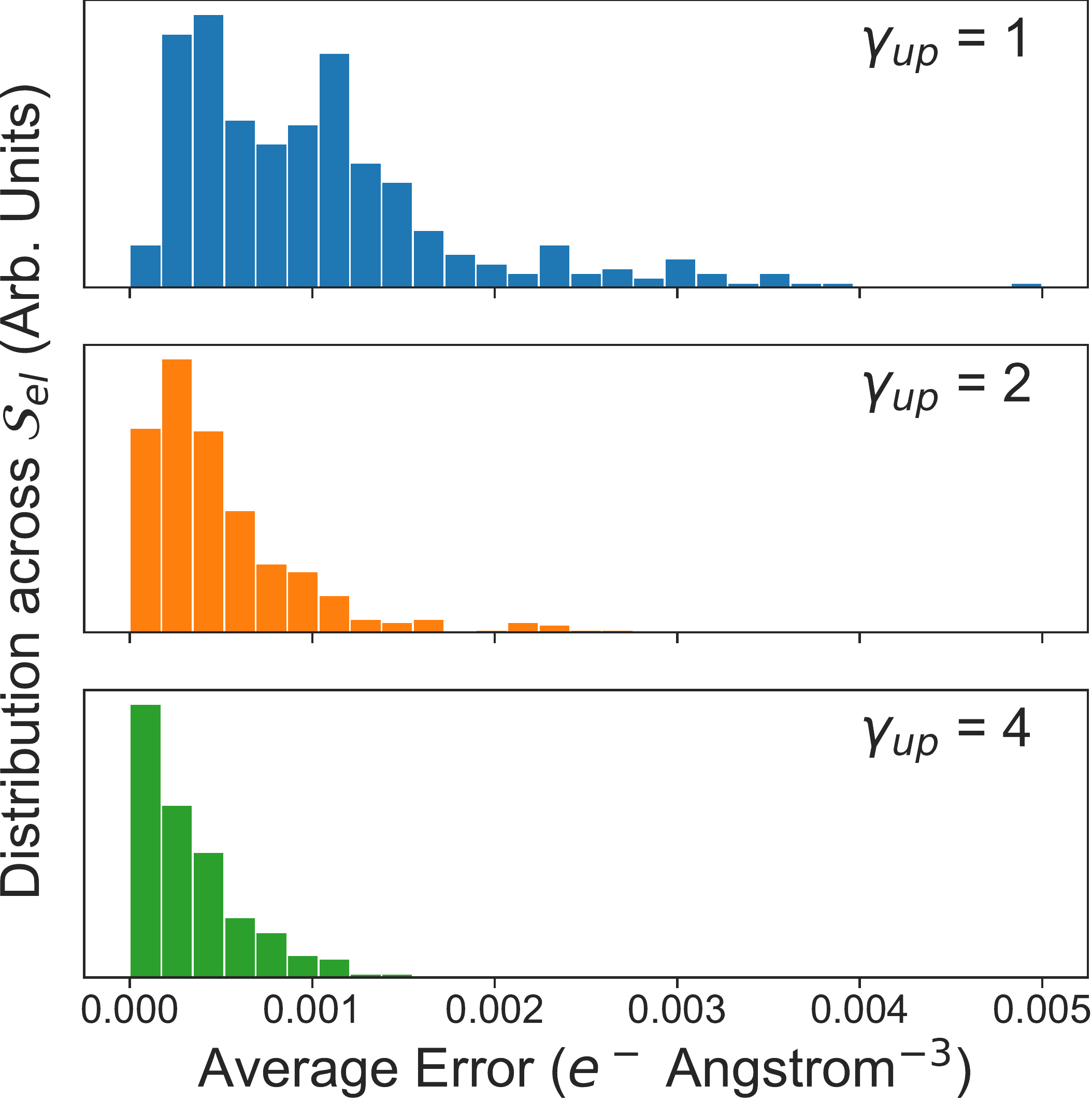}
	\caption{
	Distributions of the errors between re-sampled charge density and explicitly calculated charge densities.
	}
	\label{fig:avg_err_dist}
\end{figure}

For each charge density obtained using an explicit super-cell calculation, we obtain the average error compared to a super-cell charge density obtained from transforming the charge density.
The results of the comparison are shown in Figure~\ref{fig:avg_err_dist}.
We observe that using an up-sampling factor of 4, results in a periodic grid  fine enough such that the pseudo-charge density, which typically ranges from 0 to 100~$e^{-}/\angstrom$ near the atomic cores, only exhibits a difference of 0.001~$e^{-}/\angstrom$.

\section*{Usage Notes}\label{sec:usage}

To faciliate access to data, convenience functions have been implemented as part of the Materials Project API python client. 
These are contained within the \texttt{MPRester} class as part of the \texttt{pymatgen} software package (\url{https://github.com/materialsproject/pymatgen}). 
More specifically, two functions are provided to send independent requests to the API endpoints. 
These take as input the Materials Project ID associated with a given material in the database. 
The calculation input data from the tasks endpoint is then returned as a set of key-value pairs within a python dictionary, and the charge density data is de-serialized and returned as a \texttt{pymatgen} CHGCAR object. 
With the \texttt{MPRester} class imported, the following code workflow can be used.

% \begin{minted}[mathescape,
%               linenos,
%               numbersep=5pt,
%               gobble=2,
%               frame=lines,
%               framesep=2mm]{python}
%     mpr = MPRester(API_KEY)
    
%     # Obtain all calculation (task) IDs for a given Materials Project ID
%     task_ids = mpr.get_charge_density_calculation_ids_from_material_id("mp-149")
    
%     # Get VASP input files for a chosen calculation (task) ID
%     vasp_calc_details = mpr.get_charge_density_calculation_details(task_id)
    
%     # Obtain CHGCAR object for a given calculation (task) ID
%     chgcar = mpr.get_charge_density_from_calculation_id(task_id)
% \end{minted}

    % mpr = MPRester(API_KEY)
    
    % # Obtain all calculation (task) IDs for 
    % # a given material with a know Materials Project ID (MPID)
    % task_ids = mpr.get_chg_calc_from_mpid("mp-149")
    % 
    % 
    % # Get VASP input files for a chosen calculation (task) ID
    % chg_calc_details = mpr.get_chg_calc_details(task_id)
    % 
    % # Get the CHGCAR object for a given calculation material ID
    % 
\begin{lstlisting}[language=Python]
# Obtain the CHGCAR object for a given calculation material ID
with MPRester(<API_KEY>) as mpr:
    chgcar = mpr.get_chgcar_from_mpid("mp-149")
    
# To obtain the full list of inputs for the charge density calculation
with MPRester(<API_KEY>) as mpr:
    chgcar, calc_inputs = mpr.get_chgcar_from_mpid("mp-149",
                                            inc_inputs = True)
    
\end{lstlisting}

In order to alter the representation of the charge density obtained from the API endpoint, the \texttt{pyRho} python package (\url{https://github.com/materialsproject/pyRho}) can be used alongside the obtained \texttt{pymatgen} (\url{https://github.com/materialsproject/pymatgen}) \texttt{CHGCAR} object. Examples of how to re-grid, interpolate, and visualize are included in the repository as a set of Jupyter~\cite{Kluyver:2016aa} notebooks.

\section*{Acknowledgements}
This work was supported by the US Department of Energy, Office of Science, Office of Basic Energy Sciences, Materials Sciences and Engineering
Division under contract no. DE-AC02-05-CH11231 (Materials Project program KC23MP).

\section*{Author contributions}

JXS developed the regridding analysis software; JXS and SD developed the back-end API and JMM front-end API; JMM performed the DFT calculations that produced the charge densities; JXS, JMM, MKH, PH, and SD also participated in aggregating, ingesting and maintaining the data at different stages.
KAP was responsible for supervising and advising the project at all stages.

\section*{Competing interests}

The authors declare no competing interests.

\section*{Figures}


\newpage

\subsection*{Citing Data}
In line with emerging industry-wide standards for data citation, references to all data sets described or used in this manuscript should be cited in the text with a superscript number and listed in the ‘References’ section in the same manner as a conventional literature reference. See the examples above.

\bibliographystyle{unsrt}
\bibliography{BIBLIO}

\end{document}